\documentclass[journal=nalefd,manuscript=article]{achemso}

\usepackage[version=3]{mhchem} 

\author{Rui Pang}
\affiliation{School of Physics and Engineering, Zhengzhou University, Henan 450001, China}
\author{Bei Deng}
\affiliation{Department of Physics, South University of Science and Technology of China, Shenzhen 518055, China}
\author{Xingqiang Shi}
\affiliation{Department of Physics, South University of Science and Technology of China, Shenzhen 518055, China}
\email{shixq@sustc.edu.cn}

\title
  {Giant atomic magnetocrystalline anisotropy from degenerate orbitals around Fermi level} 

\abbreviations{Giant magnetic anisotropy energy, atomic dimer, spin-orbital coupling, defected graphene}
\AbstractOn
\begin{document}
\begin{abstract}
Nano-structures with giant magnetocrystalline anisotropy energies (MAE) are desired in designing miniaturized magnetic storage and quantum computing devices. Through ab initio and model calculations, we propose that special $p$-element dimers and single-adatom on symmetry-matched substrates possess giant atomic MAE of 72-200 meV with room temperature structural stability.  The huge MAE originates from degenerate orbitals around Fermi level. More importantly, we developed a simplified quantum mechanical model to understand the principle on how to obtain giant MAE for supported magnetic structures. These discoveries and mechanisms provide a paradigm to design giant atomic MAE in nanostructures.
\end{abstract}

Magnetocrystalline anisotropic energy (MAE), a key property of magnetic materials\cite{mae_applications}, is of crucial importance to create the energy barrier that locks the magnetic moment of the recording unit and prevents its transition to other directions. Its value is accessible for several experiments such as X-ray absorption spectra, X-ray magnetic circular dichroism and inelastic electron tunneling spectroscopy\cite{xray, biblebook,iets}. A large MAE could guarantee a long-time-stability and a thermal stability of the magnetic moment\cite{mae_application2}. For this reason, systems with large MAE attract interests of tremendous technological fields such as high-density-recording, quantum computation and molecular spintronics\cite{mae_applications,quantum_computation,spintronics1,spintronics2}.
 To maintain the direction of the magnetic moment over one year, the total MAE for a recording unit should be at least  40 $k_B$T\cite{biblebook}. However, the typical MAE of a single 3\emph{d} atom is 0.01 meV in bulk crystal and 0.1 meV in surface\cite{low_dimensional_magnetism}. Therefore, to achieve the desired large total MAE, a large magnetic cluster is required, which limits the miniaturization of the magnetic recording devices\cite{magnetic_cluster}.

In systems with lower symmetry such as single adatoms, dimers, nanowires and clusters, giant atomic MAE is more realizable.\cite{mgo_co1,mgo_co2,mgo_co3,mgo_fe,mae_nanowire,mae_adatom_mos21, mae_adatom_metal1,mae_adatom_metal2,mae_adatom_metal3,mae_adatom1,mae_adatom_CuN1,mae_dimmer1,mae_dimmer2,mae_dimmer3,mae_trimmer_graphene1,dimmer}. Among these systems, an Os adatom on MgO(100) was reported to have a record-breaking MAE of 208 meV\cite{mgo_co3}. However, the spin-flip scattering between electrons in substrates and adatoms limits the spin relaxation time, despite of a giant MAE\cite{mgo_co1,mgo_co2}. Recently, an increasing number of investigations focused on supported transition-metal-dimers, giving that specific dimers on substrates could form vertical structures.  The outermost
atom of the supported dimer could generate giant MAE as large as 100 meV and might be increased to 150 meV with the tuning of external electrical fields, which leads to a new research field of building magnetic recording devices at the molecular scale\cite{mae_dimmer_gr1,mae_dimmer_gr2,mae_dimmer_gr3,mae_dimmer_gr4,mae_dimmer_gr5,mae_dimmer_gr6,mae_dimmer_gr7,mae_dimmer_pc1,mae_dimmer_pc2,mae_dimmer_graphyne}. As the spin-orbital coupling constant $\lambda$ is proportional to Z$^4$, where Z is the atomic number\cite{biblebook}, dimers containing 5\emph{d}-elements were predicated to give the largest MAE in the $d$ electron systems\cite{mae_dimmer_gr1}. The main group late $p$-elements, which lie on the right hand side of $d$-elements in the periodic table, typically have larger $\lambda$ than the transition metals in the same period\cite{mae_radical}. Normally, these \emph{p} electron systems are not prototype systems related to magnetism so that only few organic radicals¡¯ MAE  have been paid attention\cite{mae_radical}. However, the isolated $p$ atoms such as Bi and Tl are magnetic. By designing a suitable surface ligand filed, the \emph{p}-element¡¯s magnetization can be preserved on surfaces. More interestingly, there are only six $p$ orbitals while the number of $d$ orbitals is ten, hence the coupling between orbitals associated with SOI is simpler to handle in the $p$ electron systems, which makes the design of \emph{p}-element adsorption structures much easier for the purpose of optimizing MAE. To our knowledge, no attempt has been made to construct supported magnetic structures with these $p$ elements.

\begin{figure}
\includegraphics[width=0.45\textwidth]{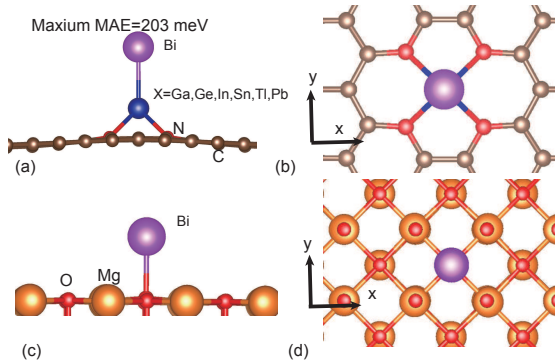}
\caption{\label{structure} Top- and side-views of Bi-X on nitrogenized divacancy of graphene (a)-(b), and of Bi on MgO(100) (c)-(d).}
\end{figure}

To address the above issue and confirm our expectation, in this letter, we construct a series of stable molecular magnetic structures with \emph{p}-elements by placing a Bi adatom or Bi-X dimmers (X = Ga, Ge, In, Sn, Tl and Pb) on symmetry matched surfaces, i.e., \emph{p}-element dimers on nitrogenized divacancies (NDV) of graphene (FIG.~\ref{structure},  the NDV has been proposed and synthesized\cite{ndv_gr1,carbon_nanotube_dv}), and \emph{p}-element adatoms on MgO(100) (Figure 1c). We discover that most of these structures have extremely large MAE (up to 200 meV) with vertical-easy-magnetization-axis, suggesting that they can be applied in magnetic storage and quantum spin processing. Based on these discoveries, we develop a quantum mechanical model and propose that in order to get giant MAE with heavy $p$ elements, one can try to create $p_{x/y}$ degenerate orbitals around Fermi level and move the $p_z$ orbital from Fermi level as far as possible. Additionally, we argue that the similar idea is applicable to $d$-and $f$-electron systems. These provide us a paradigm for atomic magnetic-storage-device design.

To elaborate the principles to generate giant $p$ electron MAE, first we focus on systems of Bi-X on NDV (Bi-X@NDV) of graphene, and then we demonstrate that this principle can be applied to other magnetic systems.

For Bi-X@NDV, by examining the total energies of a set of possible structures\cite{SI}, we find that for X = Ga, Ge, In and Sn, the vertical structure (shown in Figures~\ref{structure}a and ~\ref{structure}b) is the most stable. For X = Tl, the vertical structure is metastable but protected by an energy barrier of 3.6 eV which is sufficiently high to keep the structure at room temperature\cite{SI}. For X = Pb, the vertical structure is unstable (See Supporting Information).  Nevertheless, the properties of all the six vertical Bi-X structures will be discussed on the same footing in the following, in order to gain more insights into the principle of $p$-electron magnetism for giant MAE.

\begin{table}
\caption{\label{MAE}Spin moments (M$_S$, in $\mu_{B}$) of Bi and X, and magnetocrystalline anisotropy energies (MAE, in meV) of Bi-X on nitrogenized divacancy of graphene, positive MAE indicates a vertical easy axis.}
\begin{tabular}{ccccccc}
\hline
X           &  Ga  &  Ge    & In  & Sn    & Tl    & Pb           \\      \hline
M$_S$(Bi) & 0.85     &     0.64   &   0.85  &   0.55    &   0.77    &   0.55           \\
M$_S$(X)    &  0.02    &   -0.01     &   -0.01  &    0.00   &    -0.01   &    0.01          \\
MAE         &  -179.0    &     72.2   & 109.4    &   81.7    &    202.7   &   170.6           \\

\hline
\end{tabular}
\end{table}

We present the spin moments (M$_S$) of Bi and X alongside with the MAE of Bi-X dimers in Table~\ref{MAE}.  The MAE is defined as $\mathrm{MAE} = E_x-E_z$ where $E_{z(x)}$ is the total energy of the system when the magnetization axis is along the surface normal $z$ (along $x$ shown in Figure~\ref{structure}). The magnetization is mainly distributed in Bi atom while the atom X has nearly no M$_S$ in all these systems. The M$_S$ of Bi with 3rd group elements are 0.85, 0.85 and 0.77 $\mu_B$ for Ga, In and Tl, while those with the 4th group elements are 0.64, 0.55 and 0.55 $\mu_B$ for Ge, Sn and Pb, respectively.  More importantly, we note that all these systems have giant values of MAE together with vertical easy axis except for Bi-Ga. The vertical easy-axis is crucial for applications because it is unique and the energy difference between magnetization along in-plane directions is typically small (no more than 10$\%$ of the corresponding MAE in our cases). The MAEs of Bi-Ge, Bi-In and Bi-Sn are comparable to the largest MAE reported for $\emph{d}$-element dimers\cite{mae_dimmer_pc2}. To the knowledge of us, the MAE of the Bi-Tl dimer  of 203 meV  is the largest value ever reported for the supported dimers, and this MAE value is very close to the record-breaking value of a Os adatom on MgO\cite{mgo_co3}.
 \begin{figure}
\includegraphics[width=0.45\textwidth]{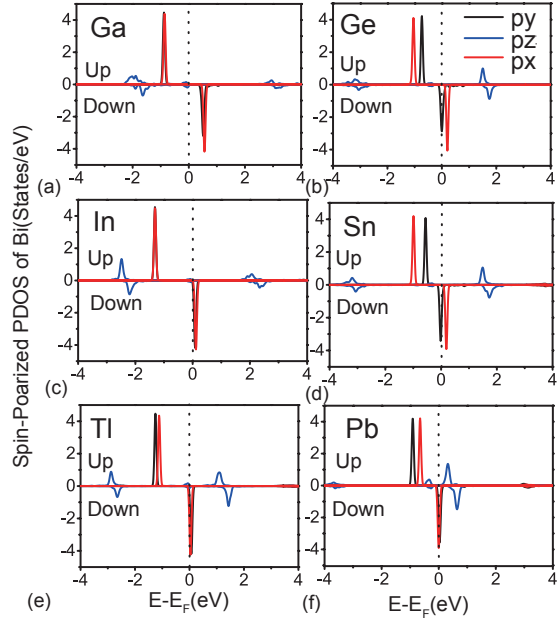}
\caption{\label{PDOS}
(a) to (f): Spin-polarized PDOS of $p$ orbitals of Bi in Bi-X@NDV (X = Ga, Ge, In, Sn, Tl and Pb) without SOI. The dashed vertical lines indicate Fermi level. }
\end{figure}

\begin{figure}
\includegraphics[width=0.45\textwidth]{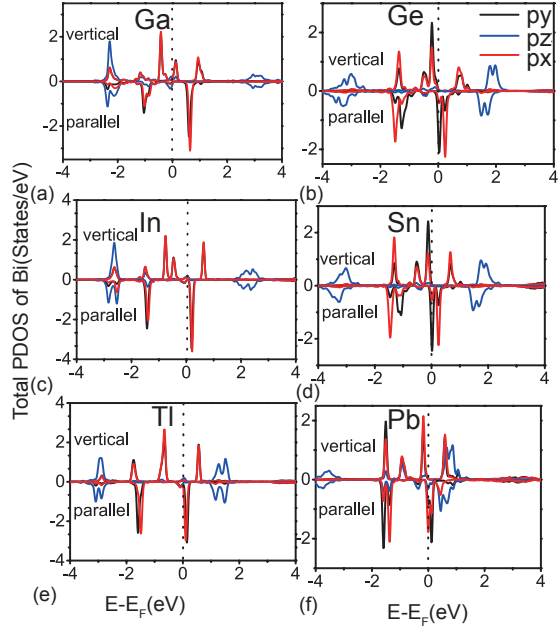}
\caption{\label{PDOSSOC}
Total PDOS of $p$ orbitals of Bi in Bi-X@NDV with SOI for magnetization along parallel- and vertical-directions. The dashed vertical lines indicate Fermi level.}
\end{figure}

 Note also from Table~\ref{MAE} that the MAE values increase along with the increasing atomic number of element X for X in the same group. However, as X is not spin-polarized, the MAEs are mainly  from Bi. In order to understand the origins of these giant MAEs as well as their dependence on the atomic number of X, in Figure~\ref{PDOS} we plot the projected density of states (PDOS) of Bi in Bi-X@NDV calculated without SOI and those with SOI in Figure~\ref{PDOSSOC}. As the $\emph{s}$ orbital ($l$ = 0) has no contribution to the SOI, only $\emph{p}$ orbitals are considered in DOS plotting.

 Figure~\ref{PDOS} shows the spin-polarized PDOS of Bi. A clear character found in these PDOS is that the \emph{p}$_x$ and \emph{p}$_y$ orbitals are quasi-degenerate. The degeneracy is associated with the approximate C$_{4v}$ symmetry of the NDV substrate in which the nearest neighbor N-N distances are 2.71 and 2.81 ${\mathrm{\AA}}$. The degeneracies of Bi bonded with the 3rd group elements are higher than the ones bonded with the 4th group elements. In the latter cases Bi gains more electrons from the X atom, which may lift the degeneracy of Bi orbitals. These degenerate orbitals are fully spin-polarized. The \emph{p}$_z$ orbital splits into two levels, one is occupied and the other is unoccupied with small spin-splitting. Except for Bi-Ga, the quasi-degenerate minority-spin \emph{p}$_{x/y}$ orbitals
lie around the Fermi level in all the other five considered systems. 

  The total PDOS of Bi with SOI are shown in Figure~\ref{PDOSSOC}. The total PDOS with magnetization in vertical ($z$ axis) and parallel ($x$ axis) directions are presented. When the magnetization is in the parallel direction, the PDOS are almost the same as the ones without SOI. However, when the magnetization is in the vertical direction, the \emph{p}$_x$ and \emph{p}$_y$ orbitals split in energy. For these orbitals just lie around the Fermi level, the energy-splitting caused PDOS change has a direct relation with MAE (see below). In contrary, the change of \emph{p}$_z$ orbitals is negligible in all cases due to $l_m$ = 0.

 Such above behaviors of the \emph{p} orbitals with SOI in Figure 3 and the origin of the giant MAE can be understood by a simplified quantum mechanical model, i.e., a combination of a two-level degenerate perturbation theory and a second-order nondegenerate perturbation theory\cite{mae_perturbation3}. The SOI Hamiltonian can be approximately written as
$\hat{V}=\lambda \mathbf{\hat{L}}\cdot \mathbf{\hat{S}}$
where $\lambda$ is an orbital and element dependent parameter containing the radial part of the orbital wave function. The angular and spin parts of \emph{p} wavefunctions can be expressed as\cite{tb_soc2}
\begin{eqnarray}
p_x(\pm)&=&\frac{Y^1_1-Y^{-1}_1}{\sqrt{2}}\chi^{\pm 1/2}\\
p_y(\pm)&=&\frac{Y^1_1+Y^{-1}_1}{\sqrt{2}i}\chi^{\pm 1/2}\\
p_z(\pm)&=&Y_1^0\chi^{\pm 1/2}
\end{eqnarray}
where $\pm$ labels the majority and minority spin, $\chi$ is the wave function in the spin space, $Y_l^m$ is the spherical harmonics. So the Hamiltonian $\hat{H}_0$ expanded in a pair of degenerated \emph{p}$_{x/y}$ orbitals of the minority spin can be expressed as a diagonalized matrix with diagonal elements being 0 and $\Delta$,
where $\Delta$ is the energy difference between the minority spin \emph{p}$_x$ and \emph{p}$_y$ orbitals around Fermi level calculated by DFT without SOI (as shown in Figure~\ref{PDOS}).

When the magnetic moment is in the vertical direction, the SOI Hamiltonian is $\lambda \hat{L}_z\hat{S}_z$, calculating the matrix elements one get $\hat{H}=\hat{H}_0+\hat{V}$
\begin{equation}\label{degenerate}
\hat{H}=\left(
\begin{array}{cc}
0&i\lambda/2\\
-i\lambda/2 &\Delta\\
\end{array}
\right)
\end{equation}
By diagonalizing this Hamiltonian, the SOI-induced new energy levels are obtained:
\begin{equation}
E_{1,2}(degenerate)=\frac{\Delta \pm \sqrt{\Delta^2+\lambda^2}}{2}\\
\end{equation}
If $\Delta=0$, $E_{1,2}=\pm \frac{\lambda}{2}$. So including SOI will lift the degeneracy of the \emph{p}$_{x/y}$ orbital pair in vertical magnetized cases. The as-split orbitals move up- and down-wards by an amplitude of $\frac{\lambda}{2}$, separately. For Bi-X (X = In, Tl and Pb), the degenerate orbitals without SOI locate at the Fermi level (as can be seen from Figure~\ref{PDOS}c, ~\ref{PDOS}e and ~\ref{PDOS}f), such lift of the degeneracy will directly lower their energy by $\frac{\lambda}{2}$. For the Bi-X (X = Ge and Sn), $\Delta\neq0$, but $\Delta$ is quite small so the above argument is still hold qualitatively. When the magnetization is along the $x$ axis, the SOI Hamiltonian is $\lambda \hat{L}_x\hat{S}_x=\lambda\frac{\hat{L}^+\hat{S}^++\hat{L}^+\hat{S}^-+\hat{L}^-\hat{S}^++\hat{L}^-\hat{S}^-}{4}$, where $\hat{L}^{\pm}$ and $\hat{S}^{\pm}$ are angular moment ladder operator and spin ladder operator. By calculating the matrix elements between $p_x(-)$and $p_y(-)$, we find that all these values are zero. So including SOI will not have much influence in energy for parallel magnetization.
Such mechanism requires the degenerate orbitals locating near the Fermi level. For Bi-Ga (see Figure 2a), the degenerate orbitals are far away from the Fermi level. The lowest unoccupied molecular orbital (LUMO) and highest occupied molecular orbital (HOMO) are at 0.65 and -0.8 eV, respectively. However, since the $\lambda$ of Bi is approximately 0.8 eV\cite{bi_constant}, the splitting by including SOI is $\lambda/2$ = 0.4 eV which is insufficient to change the occupation number. This is verified in Figure~\ref{PDOSSOC}a. Thus the degenerate orbital coupling has negligible influence on the MAE of Bi-Ga. Instead, the coupling between the nondegenerate orbitals should give the dominating contribution to the MAE in this case. In the framework of a second-order nondegenerate perturbation\cite{mae_perturbation1,mae_perturbation2}, the SOI is
\begin{equation}\label{nondegenerate}
E_{SOI}(nondegenerate)=-\lambda^2\sum_{u,o}\frac{|\langle o|\mathbf{L}\cdot \mathbf{S}|u \rangle|^2}{E_u-E_o}
\end{equation}
So up to the second-order perturbation, $E_{SOI}\leq0$.
Because the SOI is inversely proportional to the energy gap, from Figure~\ref{PDOS}a the most dominating coupling may arise from the occupied $p_z$ around -2 eV and the occupied \emph{p}$_{x/y}$  orbitals at -0.8 eV, coupling to the unoccupied \emph{p}$_{x/y}$  around 0.6 eV. When the magnetization is in the $z$ direction,  all the above mentioned matrix elements are zero; but when the magnetization is along the $x$ axis, $\langle \emph{p}_z(+)| L_xS_x|\emph{p}_y(-)\rangle=i\frac{\lambda}{2}$. Therefore, the system is more stable when being magnetized in-plane. We should mention that such effects also exist in other Bi-X cases, which makes it reasonable to interpret the variation of MAE over the element number of atom X as shown in Table~\ref{MAE}. As mentioned above, the nondegenerate part is contributed from the coupling between $p_z$ and $p_x$. In such cases, from Eq. (\ref{nondegenerate}) one can see that $E_{SOI}$ is proportional to the occupancy of $p_z$. By integrating the PDOS of $p_z$ shown in Figure~\ref{PDOS}, we find that for those X in the same group, the occupation of Bi $p_z$ orbital become smaller with the increasing of element number\cite{SI}. The reduction of $p_z$ may be attributed to the reducing of covalency in Bi-X bond which is resulted from the increasing of metallicity of X. Finally, we tired DFT+U with U = 2 eV on Bi in Bi-Tl but found negligible influence on the PDOS of Bi, which means MAE is unlikely to be dependent on the parameter U.
.



\begin{figure}
\includegraphics[width=0.45\textwidth]{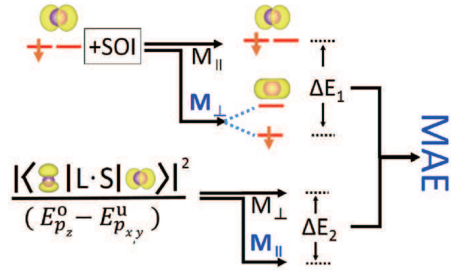}
\caption{\label{mechanism}
Illustration of the competition of two mechanisms of MAE from degenerate (upper part) and the nondegenerate (lower part) orbitals. The red lines indicate initially degenerate \emph{p}$_{x/y}$ energy levels without SOI, the orange arrow indicates spin, and the blue letters indicate the effective contributions to MAE (M$_{\perp}$ for the degenerate case and  M$_{\parallel}$ for the nondegenerate case).}
\end{figure}

The above two mechanisms are illustrated in Figure~\ref{mechanism}. In the presence of degenerate \emph{p}$_{x/y}$ orbitals, the SOI  between the $p_x$ and $p_y$ has no effect when $M_s$ is in-plane; but splits the degenerate oribtals when $M_s$ is out-of-plane (M$_{\perp}$) and hence lowers the energy of the system. For the nondegenerate orbitals, the SOI  between $p_z$ and \emph{p}$_{x/y}$ has negligible effect when $M_s$ is out-of-plane; but has significant effect on the energy lowering of the system when $M_s$ is in-plane (M$_{\parallel}$). The final results depend on the competition between these two effects from both degenerate and nondegenerate orbitals.
As $\lambda<E_u-E_o$ typically, the coupling between nondegenerate orbitals is a higher-order smaller contribution compared to the coupling between the degenerate orbitals. So usually the degenerate case dominates. We noticed that in some of the available investigations, the coupling between the degenerate orbitals were overlooked in the interpretation of MAE, in spite of the presence of the degenerate orbitals\cite{mae_dimmer_pc2,mae_dimmer_gr1,mae_dimmer_graphyne}. Our work points out the necessity to consider the effects of degenerate orbitals and makes the picture on the physical origin of giant MAE completed.

The above two mechanisms suggest a general picture to realize giant MAE. In order to generate vertical magnetization, degenerate orbitals around Fermi level are recommended. This can be obtained through manipulating an adatom or dimer on a symmetry matched substrate by electric fields or by doping the substrate. In order to reduce the in-plane MAE (associated with nondegenerate orbital coupling), the orbital splitting of $l_m$=0  need to be enlarged. As atom X lies above the graphene surface, the Bi atom which is responsible for the giant MAE has a larger distance from the substrate than an adaom has in adatom/MgO\cite{mgo_co1}. This implies that the influence of the surface spin-flip scattering will be smaller in our cases so the system we proposed is hopeful to have a longer spin relaxation time. Note that in a certain crystal field, the splitting of degenerate orbitals under SOI could also take place for $d$ and $f$ orbitals. The orbital of $l_m$=0 always contributes to in-plane MAE. Thus these conclusions should also hold for a broad range of magnetic systems.
To further demonstrate the mechanisms shown in Figures~\ref{mechanism} are generally applicable, we investigate a single Bi adatom on MgO(100). The most stable geometric structure is subject to the atop adsorption on oxygen (Figure~\ref{structure}c and ~\ref{structure}d), and accordingly the quasi-degenerate orbitals appeared around Fermi-level with an MAE being 75 meV, which again suggests the proposed mechanisms hold in other analogical  systems\cite{SI}.

In summary, we discovered that most of the vertical structures of Bi-X@NDV of graphene can be stable at room temperature, and  Bi is spin-polarized in these systems. Most of these dimers show giant MAE with easy axis in the out-of-plain direction. In particular, the MAE of Bi-Tl@NDV is comparable to the largest MAE ever reported. The origin of giant MAE can be attributed to the lift of the degeneracy with SOI of the orbitals near the Fermi level without SOI. Such degeneracy is relevant to the local C$_{4v}$ symmetry of the system. These finding provides a novel and general method to design giant MAE with \emph{p}- ,\emph{d}- and even \emph{f} electron magnetism: one could find a substrate with a proper symmetry (C$_{4v}$ for $p$ in this paper) to create quasi-degenerate orbitals and then tune these orbitals close to Fermi level; and simultaneously move the orbitals with $l_m$=0 far from Fermi level. These discoveries and insights pave a pathway for further development of robust nanomagnetic units.

\begin{acknowledgement}
This work is supported by the National Natural Science Foundation of China (Grants 11474145 and 11334003), the Young Teachers Special Startup Funds of Zhengzhou University, and the Special Program for Applied Research on Super Computation of the NSFC-Guangdong Joint Fund (the second phase).

\end{acknowledgement}

\begin{suppinfo}

This file contains the discussion on the computational detail, stability of structures, p orbital occupacies and geometrical information of the structures investigated.

\end{suppinfo}



\begin{tocentry}
\includegraphics[width=0.8\textwidth]{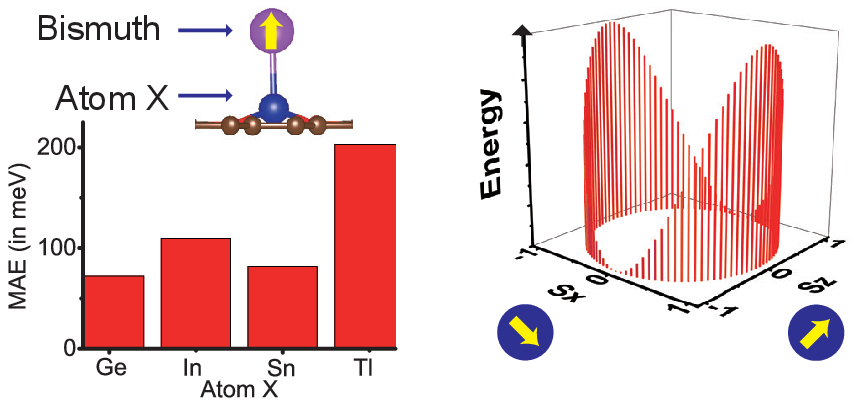}

\end{tocentry}
\bibliography{mybib}
\end{document}